\documentclass[aps,prl,twocolumn,groupedaddress]{revtex4}
\newcommand{\nmg}{Ni$_{2}$MnGa}

\usepackage[dvips]{graphicx}
\usepackage{amssymb}

\newcommand{\Ltw}{$\mathrm{L2_1}$}
\newcommand{\TA}[1]{$\mathrm{TA_{#1}}$}
\newcommand{\LA}{$\mathrm{LA}$}

\begin{document}

\title{A First-Principles Investigation of Phonon Softenings and Lattice
  Instabilities in the Shape-Memory System Ni$_{2}$MnGa} 

\author{A. T. Zayak, and P. Entel}
\affiliation{Institute of Physics, Gerhard-Mercator University,
             47048, Duisburg, Germany}
\author{J. Enkovaara, A. Ayuela, and R. M.  Nieminen}
\affiliation{Laboratory of Physics, Helsinki University of Technology,
             02015 Espoo, Finland }

\date{\today}


\begin{abstract}

Ferromagnetic \nmg\ has unique magnetoelastic
properties. These are investigated by detailed computational studies
of the 
phonon dispersion curves
for the non-modulated cubic \Ltw\ and tetragonal structures.
For the \Ltw\ structure, a complete softening of the transverse acoustic mode
has been found around the wave vector $\mathbf{q}=[1/3,1/3,0](2 \pi/a)$.   
The softening of this \TA{2}\ phonon mode leads to the premartensitic modulated
super-structure observed experimentally. Further phonon anomalies, related to other structural
transformations in \nmg, have also been found and examined.
These anomalies appear to be due to the coupling of 
particular acoustic phonon modes and optical modes derived from Ni. 

\end{abstract}

\pacs{}

\maketitle


Displacive, diffusionless structural transformations of the martensitic type
are known to occur in many metallic alloys.
These transformations involve cooperative rather than diffusive
displacements of atoms and are often associated with phonon anomalies in
the parent phase and related precursor phenomena. 
It is a general challenge of
the fundamental physics to explain driving forces of the martensitic
transformations.

Since the discovery of a martensitic transformation in ferromagnetic \nmg\
\cite{Webster},
this material has attracted strong interest \cite{Nature2}. 
This Heusler alloy is one of the very rare
magnetic  materials which undergo a martensitic 
transformation below the Curie temperature, whereby the combination of
magnetic and structural features is responsible for its unique
magnetomechanical properties.  Shear deformations of more than 5 \% have been
obtained in  magnetic fields 
\cite{Ullakko2,Murray,Sozinov}. 
These features together make \nmg\ very efficient
for  magnetic-shape-memory (MSM) technology \cite{Ullakko}. The
MSM technology is based on the magnetic field induced redistribution of
martensitic domains in the sample. From the technological point of view, \nmg\ is  much more
promising than other materials being presently in commercial use,
 for example the well-known material Tb-Dy-Fe (Terfenol-D) which exhibits
magnetostrictive strains of about 0.1 \%.
Design of new efficient MSM
magneto-mechanical actuator devices is in progress \cite{Aaltio}.  
Also the search for new materials with magnetic shape-memory effects
is underway, even in antiferromagnets \cite{Nature1}.

\begin{figure}[h]
\resizebox{7.5cm}{!}{\includegraphics{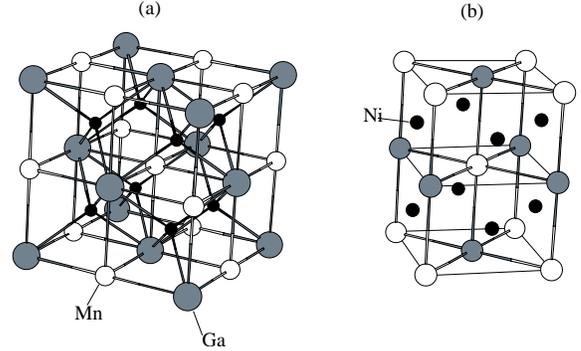}}
\caption{ \label{cub} (a) Simple cubic cell of \nmg\ (Heusler \Ltw\
  structure);  (b) tetragonal bct cell in the [110] direction.
The equilibrium lattice parameter obtained  is $a_{\mathrm{cubic}}=5.8067 $
\AA\ and agrees with the experimental value \protect\cite{Webster}.
The direction [010]$_{\mathrm{orth}}$ of the supercell corresponds
to the direction [110]$_{\mathrm{cubic}}$ of the cubic structure.
As the supercell consists of five tetragonal cells,
the lattice parameters of the supercell are related to the lattice parameters of
the cubic \Ltw\ structure as follows: 
$a_{\mathrm{orth}}=(1/\sqrt{2})a_{\mathrm{cubic}}$, 
$b_{\mathrm{orth}}=(5/\sqrt{2})a_{\mathrm{cubic}}$,
$c_{\mathrm{orth}}=a_{\mathrm{cubic}}$.}
\end{figure}

Despite the experimental and technological success, a microscopic theory is
missing, which would be able to show how far the premartensitic phase 
transition will result from the coupling of soft phonon modes and homogeneous 
strains associated with the shear constant $C'$ 
\cite{Gooding,Brown}. A complete understanding requires the evaluation of the
phonon spectrum. 
The aim of this letter is to present supercell phonon calculations for the
\Ltw\ and $\mathrm{T}$ structures (see Fig. \ref{cub} and Table \ref{tab1})
using a state-of-the-art first-principles method based on density-functional
theory.

 \nmg\ is ferromagnetic
at room temperature ($T_\mathrm{C} \approx 380$ K)  and undergoes a 
two-step martensitic phase transformation. 
As a matter of fact, a number of different thermal and stress-induced
martensitic structures have been observed in \nmg\ \cite{Mart-92}. 
Table \ref{tab1} gives a summary of the structures found for \nmg\
\cite{Webster,Brown,Mart-92}. 
In the parent phase, several phonon anomalies can be observed with decreasing
temperature. At the temperature $T \approx 260$ K,  
a nearly complete softening of the $[1/3,1/3,0] (2\pi/a)$ 
transverse acoustic \TA{2} mode with polarization along 
[1\=10] leads to a premartensitic  phase transition, which has been evidenced
by neutron scattering \cite{Zhel95,Zhel,Stuhr,Brown}, 
X-ray \cite{Fritsch}, electron microscopy \cite{Cesari}, and
ultrasonic measurements \cite{Worgull,Manosa97}
or a combination of the previous methods \cite{Manosa}. This 
structural transformation involves a commensurate periodic distortion of the
parent phase with a propagation vector equal to that of the soft mode
$\mathbf{q}=[1/3,1/3,0](2\pi/a)$ corresponding to  six atomic planes or three
lattice spacings.  
The precursor phenomena involves the magneto-elastic coupling as it has been
described using Landau-type models \cite{Planes}.
These phenomenological models take as input very important 
lattice-dynamical properties. Thus, explicit calculation of the lattice
dynamics is required since in transition metals, the {\it d}-electrons allow
for many types of coupling, and several modes can be involved.

\begin{table}[h]
 \caption{\label{tab1} Structural parameters of the  crystal structures of
Ni-Mn-Ga alloys according to Ref. \cite{Mart-92} and this work. They follow
(from left to right)  the order of possible appearance when lowering 
the temperature. There is a superimposed modulation specified by the
notation $\mathrm{nL}$, where n stands for the close commensurate number of
bct cells in the [110] direction; their respective $\mathrm{q}$ vectors are
in units of ($2\pi/a$). The so-called premartensitic transformation
happens between the \Ltw\ and $\mathrm{3L}$ structures.}
\vspace{2mm}
   \begin{tabular*}{0.48\textwidth}{@{\extracolsep{\fill}}l c c c c c }
    \hline
\hline
    Axes  &  \multicolumn{5}{ c }{Lattice parameters (\AA)} \\ 
    \hline
 & \Ltw\ & $\mathrm{3L}$ & $\mathrm{5L}$ & $\mathrm{7L}$ & $\mathrm{T}$ \\
       $a$  & 5.824 & 5.824 & 5.90  & 6.12  &  6.44 \\ 
    $b$  & 5.824 & 5.824 & 5.90  & 5.78  &  5.52 \\ 
    $c$  & 5.824 & 5.824 & 5.54  & 5.54  &  5.52 \\ 
    Modulation & none & $\mathrm{q} \approx 0.33$ & $\mathrm{q} \approx 0.43$
    & ? & none \\ 
    Tetragonality & 1 & 1 & 0.94 & - & $\approx$ 1.2 \\
\hline
   \end{tabular*}
\end{table}


For the sake of simplicity, we present phonon calculations for
the direction [110], which is the most interesting one as seen from
the measurements for the acoustic modes \cite{Zhel}. 
We have used the 
 direct method for the evaluation of the phonon dispersion curves
 \cite{Parlinski,Phonon}, whereby the
 Hellmann-Feynman forces are calculated with the
 Vienna {\it Ab-initio} Simulation Package (VASP)
\cite{Kresse-96,Kresse-99PAW} and the
implemented  projected augmented wave formalism (PAW) \cite{PAW}.
Within density-functional theory, 
the electronic exchange and correlation are treated by using the generalized
gradient approximation (GGA). The 3{\it d} electrons of Ga have 
been included as valence states. 
The importance of using PAW for \nmg\ instead of
pseudopotentials has been pointed out earlier \cite{Ayuela-priv}.
An orthorhombic supercell of 40 atoms formed by five tetragonal
crystallographic unit cells, as seen in Fig. \ref{cub}(b), has been used to
calculate the phonon spectrum for the [110] direction \cite{detail1}. 
The direct method implies that we calculate forces induced on all atoms
of the supercell when a single atom is displaced from its equilibrium 
position. Displacement of the atoms in only one tetragonal unit cell 
along all directions allows one to derive the force
constant matrix and the dynamical matrix. Diagonalization of the dynamical
matrix then leads to a set of eigenvalues for the phonon frequencies and
corresponding  
eigenvectors. The calculations have been done by using an amplitude for
the displacements of  
$u = 0.03$ \AA. This value is sufficient to calculate the forces with required
precision and to fulfill the conditions of the harmonic approximation.

\begin{figure}[h]
  \resizebox{7.5cm}{!}{\includegraphics*{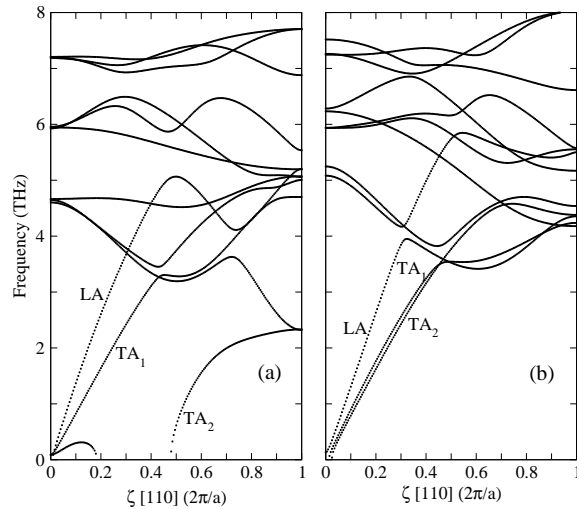}}
  \caption{ \label{disp} Phonon dispersion curves for the cubic \Ltw\
  structure (a) and the tetragonal $\mathrm{T}$ structure (b) of \nmg. Here,
   the reduced    wave vector coordinate $\zeta$ spans the 
   fcc BZ from $\Gamma$ to $X$. 
}
\end{figure}

Phonon dispersion curves
calculated for \nmg\ in the [110]$_{\mathrm{cubic}}$ direction are shown in 
Fig. \ref{disp}(a). The acoustic modes are qualitatively and quantitatively 
in  very good agreement with experimental results obtained from inelastic
neutron scattering \cite{Zhel}. For instance, following the
acoustic branch, the values of the \TA{2}\ branch at $\zeta=1$, of the
 crossing between
\LA\ and \TA{1}\ branch, and of the maximum of the \LA\ branch are
 2.33, 4.43 and 5.13 THz, in good agreement with the experimental ones, 2.67,
 4.35 and 5.07 THz \cite{Zhel}. The initial slopes of the
curves in the longitudinal ($v_\mathrm{L} = 5.23 \times 10^5 \;
\mathrm{m/s}$) and 
transverse ($v_{\mathrm{TA_1}} = 3.35 \times 10^5 \; \mathrm{m/s}$;
$v_{\mathrm{TA_2}} = 1.02 \times 10^5 \; \mathrm{m/s}$) modes agree well
with the sound velocities measured via the 
neutron scattering dispersion curves \cite{Zhel}.  
The slope of the phonon dispersion curves around the $\Gamma$ point is 
positive, which is consistent with the stability around this phase. 

We find a complete softening of the transverse acoustic \TA{2}\ phonon mode
between $\zeta =0.2$  and $\zeta =0.55$.
This softening occurs around the wave vector coordinate  $\zeta =
1/3$, which corresponds to the soft-mode phonon anomaly 
observed in the experimental studies \cite{Zhel}. 
We point out  that the experimentally observed 
premartensitic transition is related to this soft mode.

\begin{figure}[h]
\resizebox{6cm}{!}{\includegraphics*{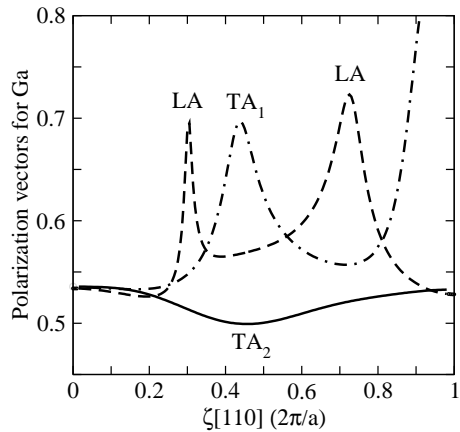}}
  \caption{\label{norm}  Polarization 
   vectors corresponding to the acoustic phonon modes of Ga. The peaks at
   $\zeta=0.300$, $\zeta=0.433$ and $\zeta=0.715$ are due to the coupling to
   the optical modes of Ni.  
}  
\end{figure}

\begin{figure}[h]
  \resizebox{3.75cm}{!}{\includegraphics*{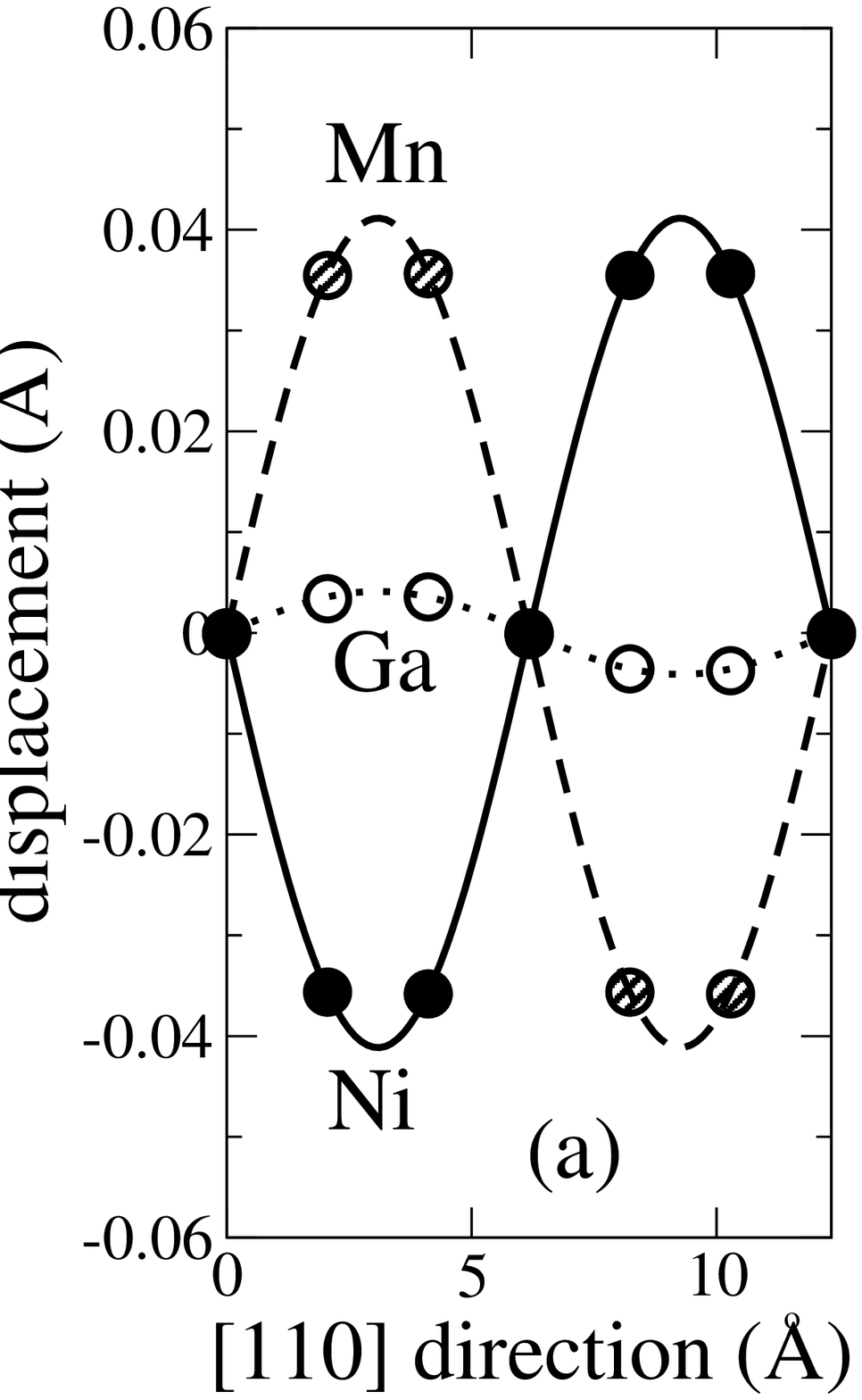}}
  \resizebox{3.75cm}{!}{\includegraphics*{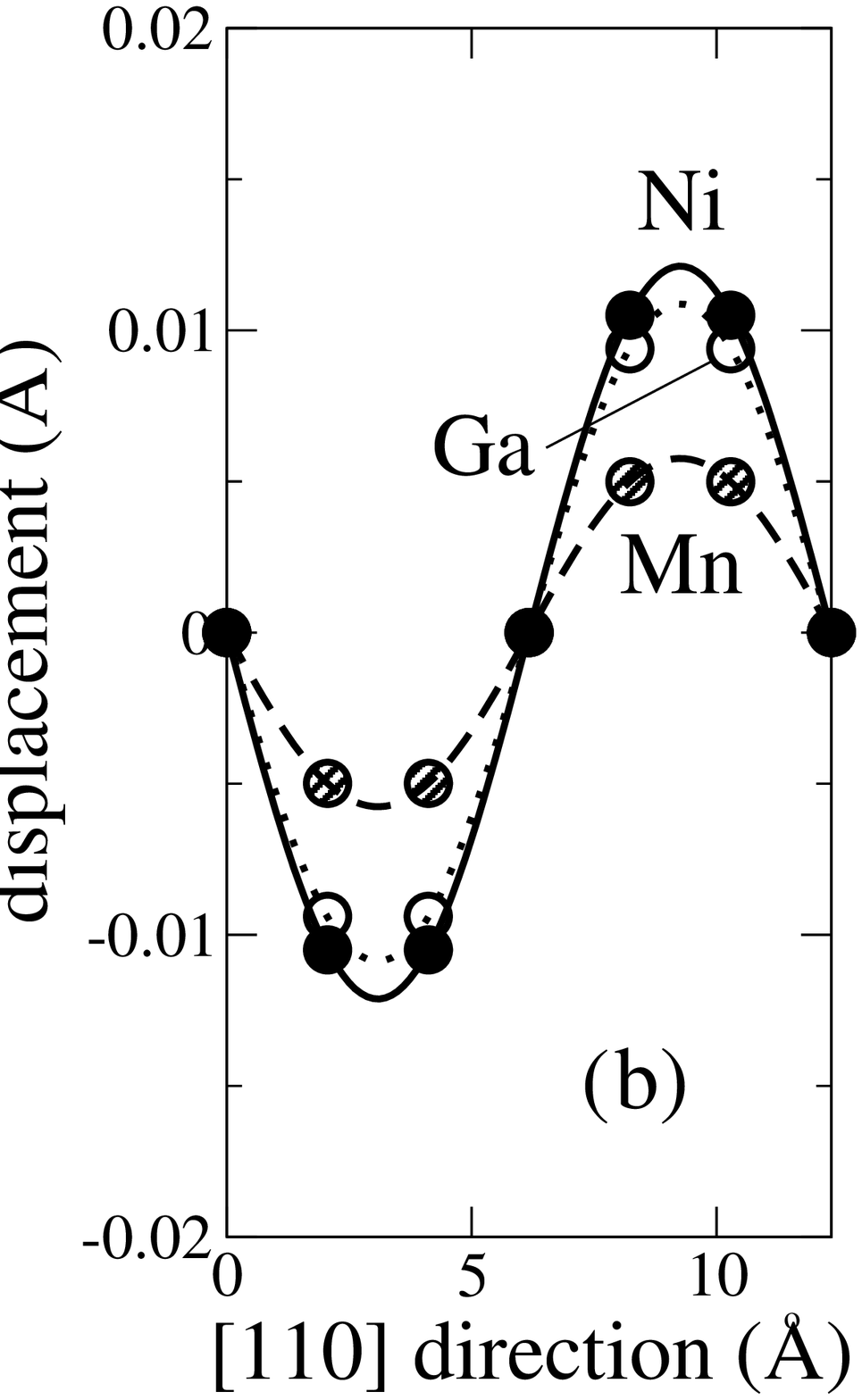}}
  \caption{\label{shu} Displacements of the $\mathrm{3L}$
premartensitic modulated structure:  before a) and after b) structural
   relaxation. The atoms are displaced parallel to $a_{\mathrm{orth}}$ in 
successive (010)$_{\mathrm{orth}}$ planes. The displacements are in a
   different scale in order to be observed.} 
\end{figure}

The softening of the \TA{2}\ mode means that the cubic \Ltw\ high-temperature
structure is unstable at zero temperature with respect to a particular 
atomic displacement leading to the
formation of a modulated premartensitic structure. 
The $\mathrm{3L}$ structure can be modeled by
a large supercell composed of three basic  bct supercells in the [110]
direction  (see Fig. \ref{shu}) \cite{detail2}.
The atoms of the input  structure  have a modulated amplitude with a maximum 
around  0.05 \AA \,  where the Ni atoms are in opposite phase to the Mn and Ga
atoms. Full relaxation of the three lattice parameters, $a$, $b$ and $c$ has
been  allowed.  Also, lower magnetizations
of 60-80 \% of the full magnetization value for 
the whole supercell has been taken into account.
In addition, two periods commensurate with
the lattice were tested, but the structure with one period, as shown
in Fig. \ref{shu}, gives the  minimum energy.
The static displacements of the $\mathrm{3L}$ modulated structure are shown  
in Fig. \ref{shu}(b). The modulation of the atoms stays in 
phase which differs from the input data, 
but it is typical for an acoustic branch such as \TA{2}. 
The amplitudes for all the atoms are nearly the same, 
although they differ slightly but not according to the atom masses. Rather,
the order is according to the sequence of the optical phonons energies, where
Ni has the lowest energy, optical modes of Ga are in the next energy windows,
and Mn optical phonons are of the highest energy (see Fig. \ref{disp}).

We have considered so far only the acoustic mode \TA{2}, 
because this mode is responsible 
for the phase transformation \Ltw\ $\rightarrow$ $\mathrm{3L}$. However, (see 
Fig. \ref{disp}) the  other two  
acoustic modes, \TA{1}\ and \LA, are coupled to the low - lying optical modes
of Ga.  Fig. \ref{norm} shows how this coupling affects the behaviour of  the
polarization vectors of Ga for its acoustic modes.
The Ga atom type have been chosen for clarity, while acoustic modes of the 
other atoms show similar trends.
Figure \ref{norm} illustrates that the acoustic-optical
hybridization (compare with Fig. \ref{disp}) leads to a number of
anomalies, which occur at the corresponding wave vector 
coordinates $\zeta = 0.3$, $\zeta = 0.43$ and $\zeta = 0.715$. 
A longitudinal-phonon optical mode can be associated to a charge density
wave, which gives a sharp peak in the magnetic susceptibility $\chi(q)$.
The  crossing at $\zeta = 0.43$  compares well with those for which anomalies
are observed in the susceptibility calculations, which have been
associated with the lattice instabilities in the $\mathrm{5L}$ and
$\mathrm{7L}$ structures  \cite{Velikokhatni,Harmon}.  
These crossings correlate in a natural way with the change 
of the $C_{44}$ elastic constant, which is, as it turns out, not related to
the \TA{2}\ branch through the premartensitic transformation. 
It comes as a novelty that  the involvement of the optical modes is required
in order to understand the phase transformations in these alloys.

 It is interesting to note that the anomaly in the \TA{2}\ branch 
has been found around the soft mode 
$[1/3,1/3,0] (2\pi/a)$ in our calculations, which is the wave vector
associated with the phonon 
anomaly of the premartensitic transition \cite{Zhel}. 
Also the phonon branches on the [110] direction for the $\mathrm{T}$ structure
are given in Fig. 2(b). In this latter case, the branches \TA{2}\ and \TA{1}\ 
become similar and the softening of the \TA{2}\ branch, present in the cubic
phase, disappears. This finding is consistent with the stability of the
$\mathrm{T}$ structure  \cite{Ayuela} and adds insight into interpreting the 
experiments \cite{Manosa}. 
The sequence of the optical modes is different in this case and they are not
degenerate at the $\Gamma$ and $X$ points.  
The instability of the cubic structure at zero
temperature is  the clue to understand how the martensite phase may nucleate. 
Instead of twinnings and stacking faults, the modulation of the martensitic
tetragonal structure of \nmg\ corresponds to a smaller structural change
easier to accommodate  or to derive from the parent cubic structure. With
respect to the electronic  structure, the lowering of the symmetry allows for
a splitting of the density of states peak 
at the Fermi level \cite{Brown,Ayuela}, thereby decreasing the valence band
energy contribution to the total energy. Thus, structural modulation and
lowering of the electronic energy explains  the 5L structure of \nmg\
\cite{Za}.      

On the other hand,  the recent susceptibility calculations
\cite{Velikokhatni,Harmon} have shown that the peak in the generalized
susceptibility  at ${\mathbf{q}} = [\zeta,\zeta,0](2\pi/a)$ for $\zeta =
0.33$ corresponds to a magnetization of about 60\%. This suceptibility 
peak moves toward higher values of $\zeta$ as the magnetization increases:
full magnetization leads to a peak at $\zeta = 0.4$.
However our calculations show no need for the extra magnetization 
hypothesis, since all the interesting $\mathbf{q}$ vectors are already
present.  This casts serious doubts on the role of magnetization in these
intermediate or premartensitic transitions. 
Secondly, this finding reinforces  the arguments that
the elastic entropy term drives the transition \cite{apl} and the 
spin spirals do not show anomalies involved in the shuffling of the
\Ltw\ structure \cite{prb}.  In view of these results,
the intermediate and the martensitic transition seem to be really independent,
 although, of course, the \Ltw\ structure is necessary in order to
 have the  intermediate transformation at $\zeta = 0.33$.


In summary, we have calculated the phonon dispersion for two 
structures of \nmg: the cubic \Ltw\ and the tetragonal $\mathrm{T}$ at zero
temperature.  The cubic structure is found to be unstable with respect to a
specific re-arrangement (shuffling) of the atoms which leads to a modulated
super-structure $\mathrm{3L}$.  We have calculated stability of the
$\mathrm{3L}$ 
structure with high accuracy and found that it reflects perfectly our phonon
calculations.  Furthermore, we have shown that the subtle competition of
low-lying optical modes of Ni and acoustic modes account for the
experimentally observed wave vectors, for which anomalies have been observed
($\mathrm{3L}$, $\mathrm{5L}$ and $\mathrm{7L}$ structures). These calculations
have been done by using first-principles methods, and are in
very good agreement with experimental findings.

\begin{acknowledgments}

This work has been supported by the Graduate School ``Structure and
Dynamics of Heterogeneous Systems'' of the Deutsche Forschungsgemeinschaft
(DFG) and by the Academy of Finland through its Centers of Excellence Program
(2000-2005). Computer facilities 
of the Research Centre J\"{u}lich (Project ''Lattice
dynamics'') are acknowledged. We thank Prof. M. Acet, Prof. K. Parlinski,
Dr. A. Postnikov and Prof. A. Planes for valuable discussions. 

\end{acknowledgments}

\newpage


\begin{thebibliography}{99}
\bibitem{detail1}
The plane-wave cutoff energy of 353.4 eV and
 the Monkhorst-Pack k-point generation scheme is used with a grid of \ 10
 $\times$ 8 $\times$ 2\ points  in the full Brillouin zone.
\end{thebibliography}
\end{document}